# Data-driven Model Predictive Control Method for DFIG-based Wind Farm to Provide Primary Frequency Regulation Service

Zizhen Guo, Wenchuan Wu, *Fellow, IEEE*

***Abstract*—As wind power penetration increases, the wind farms are required by newly released grid codes to provide frequency regulation service. The most critical challenge is how to formulate the dynamic model of wind farm for dynamic control, since it is essentially is nonlinear and there are huge amount of parameters to be maintained frequently. This paper proposes a data-driven model predictive control (data-driven MPC) method to make wind farms participate primary frequency regulation. In this method, a specialized dynamic mode decomposition (SDMD) algorithm is developed, which can linearly approximate the dynamics of wind farm from measurements based on Koopman operator theory. Compared with the existing extended dynamic mode decomposition (EDMD) method, this tailored SDMD has two advantages: 1) fully capturing the nonlinear transients of wind turbine dynamics with good accuracy under a wide range of working conditions; 2) much less computational burden with model dimensionality reduction. Based on the recursively updated linear dynamic model, a model predictive control solution is implemented. The simulation results show this model-free solution can dynamically optimize wind turbine generators' active power to track the frequency response requirement from system operator and minimize the rotor speed distortion.**

## I. Introduction

WIND energy has faced nearly a decade of rapid but rather steady growth with the global concern over climate change issues and energy crisis. According to the report from the Global Wind Energy Council (GWEC), in 2019, the new installation of global wind power is 60.4 GW, which is a 19% growth compared to 2018 and makes the total capacity reach 650 GW [1]. With the expansion of wind energy market, variable speed wind turbine generators (VSWTGs), especially doubly fed induction generators (DFIGs), are more favorable type because of their ability to track the wind speed change to optimize the wind power extraction. However, the interfacing power electronic converters between DFIGs and the utility grid have the inertia of wind turbines decoupled from the system, which results in a low-inertia system with its frequency vulnerable to load change or equipment failure [2]. The most widely adopted control mode of DFIG is maximum power point tracking (MPPT). The behavior of DFIG under MPPT mode is more like a negative load for its output power is governed by the random change of local wind speed rather than the control requirement from the whole system like global safety, stability and optimality [3].

To reinclude wind power into system-wide control to maintain system reliability, the modern grid codes of different countries or regions have published specific requirements on wind farms to provide ancillary services, which used to be counted on conventional thermal power plant [4]. As is focused in this article, the ability to provide primary frequency regulation should be addressed by wind farms before their integration into the power grid. The most common requirement is speed-droop characteristic, under which the active power of wind farm is responsive to the system frequency deviation with a droop ratio.

The present control strategies of wind power frequency support can be classified into two categories: wind turbine level and wind farm level. On wind turbine level, the local controllers of DFIG are modified to include frequency deviation term into the control loop of active power output to supply primary frequency regulation service [5]-[10]. In [5], the hidden kinetic energy of turbine blade is exploited to emulate inertia and support system frequency. However, the recovery process of rotor speed drop may lead to secondary frequency excursion. Thus, de-loading control mode is carried out by rotor over-speeding [6][7] or blade pitch control [8] to reserve active power control margin for frequency response. The authors in [9] proposed a variable droop control strategy that smoothes power fluctuation of DFIGs during frequency response process by adjusting droop ratio continuously according to the changes of wind speed. In [10], an adaptive gain control method is developed for stable inertial control of a DFIG-based wind farm. Adaptive gains are set to be proportional to the kinetic energy stored in DFIGs to avoid over-deceleration under low wind speed. On wind turbine level, local controllers operate independently without interaction between neighbor generation units, therefore the optimality of the whole wind farm operation

Manuscript received XX, 2020. This work was supported in part by the National Key Research and Development Plan of China (Grant.2016YFB0900403)) and in part by the National Science Foundation of China (51477083).

The authors are with the Department of Electrical Engineering, Tsinghua University, Beijing 100084, China (Corresponding Author: Wenchuan Wu, email: wuwench@tsinghua.edu.cn).

cannot be guaranteed.

On wind farm level, the coordination between wind turbines are explored through wind farm wide optimization control [11]-[15]. While wind farm providing droop frequency response as a single machine, the power allocation within needs to be optimized to stabilize the operation of each wind turbine and to minimize the side effect of frequency control, e.g., rotor speed distortion and mechanical load increase. In [11], a power deviation control method is developed for wind farm frequency regulation. The power reserve is allocated to each wind turbine sequentially in the order of their available power. The authors in [12] explored the impact of wake effect on the performance of wind farm frequency control. The authors in [13] give a coordinative control strategy of virtual inertia and primary frequency regulation of DFIG-based wind farm. In [14] and [15], a detailed dynamic model of VSWTG is adopted in the wind farm wide frequency control strategy, which solves the dynamic optimization control problem and dispatches the reference for each VSWTG online to ensure the stable and optimal operation of the wind farm. Because the dynamic procedure of wind power conversion features strong nonlinearity, the effectiveness of dynamic optimal control strategy of wind farm relies on the accurate and fast solution of a nonlinear optimization problem, which is hard to balance due to the large problem scale for wind farms with many generation units and the limited response time required by the system frequency control policy.

In light of this dilemma, we propose a specialized dynamic mode decomposition (SDMD) method based on Koopman operator theory [16] and its related extended dynamic mode decomposition (EDMD) algorithm [17], in the purpose of fitting a numerical hyper-dimensional dynamic model for DFIG, which is globally linearized to maintain a balance between the accuracy and the simplicity of the optimal control problem. The key idea of Koopman operator is that for a given nonlinear dynamic system, the evolution of the state-space can be globally linearized under the properly selected observable functions. Unlike local linearization method, since the observable functions are designed to compensate for the specific nonlinear features of the dynamic system, Koopman operator can govern the evolution of a hyper-dimensional observable space linearly and accurately for whole range of state variables, which is referred to as *global linearization* ability. The authors in [18] introduce Koopman operator into controlled dynamic system by viewing the product of original state-space and all control sequences as an extended state-space, which makes it possible to transfer a nonlinear model predictive control (MPC) problem to a hyper-dimensional linear MPC problem.

On the effort of exploiting the global linearization potential of Koopman operator, the choice of observable functions is an important yet still open problem. The main challenge is to build an invariant subspace including the original state for optimal control strategy. However, when we focus on the dynamic modelling of DFIG, there is prior information about the major nonlinearity that governs the major dynamic process of DFIG. Take advantage of this, we propose SDMD method which specializes the choice of observable functions for Koopman operator based on the experience-based model of the controlled object. The method inherits the merit from data-driven strategy that only input and output data is required to build a dynamic model of the controlled object. But with experience-based model linking the gap between the mathematical tool and the engineering control problem, the data-driven modelling method is powered with the human understanding of the intrinsic dynamics of the studied system, which has the potential to capture the intrinsic nonlinearity of dynamic system without a big number of randomized basis functions as observables, which is verified a leap in both accuracy and simplicity during data-driven modelling process in case study. Combining SDMD with MPC in the wind farm frequency control problem, we build a data-driven linear model of DFIG and cooperate the operation of DFIGs within the wind farm based on the frequency regulation requirements from the system operator outside and the local states, e.g., rotor speed and wind condition, inside the wind farm to ensure the safe and economic frequency control.

The remainder of this paper is organized as follows. Section II describes the experience-based dynamic model of DFIG. Section III presents the main tool, Koopman operator and the proposed SDMD method for DFIG modelling and control. The data-driven MPC method for wind farm primary frequency control is illustrated as well. A case study of single DFIG that demonstrates the advantage of SDMD modelling method and a case study of wind farm integrated 4-bus system that demonstrates the effectiveness of proposed frequency control method are both presented in section IV. Finally, section V draws the conclusion.

## II. Wind Turbine Dynamic Model

In this paper, we consider the modelling and control of DFIG type wind power conversion system due to its wide range of installation and promising market share. A schematic diagram of main components and their connection in DFIG is shown in Fig. 1.

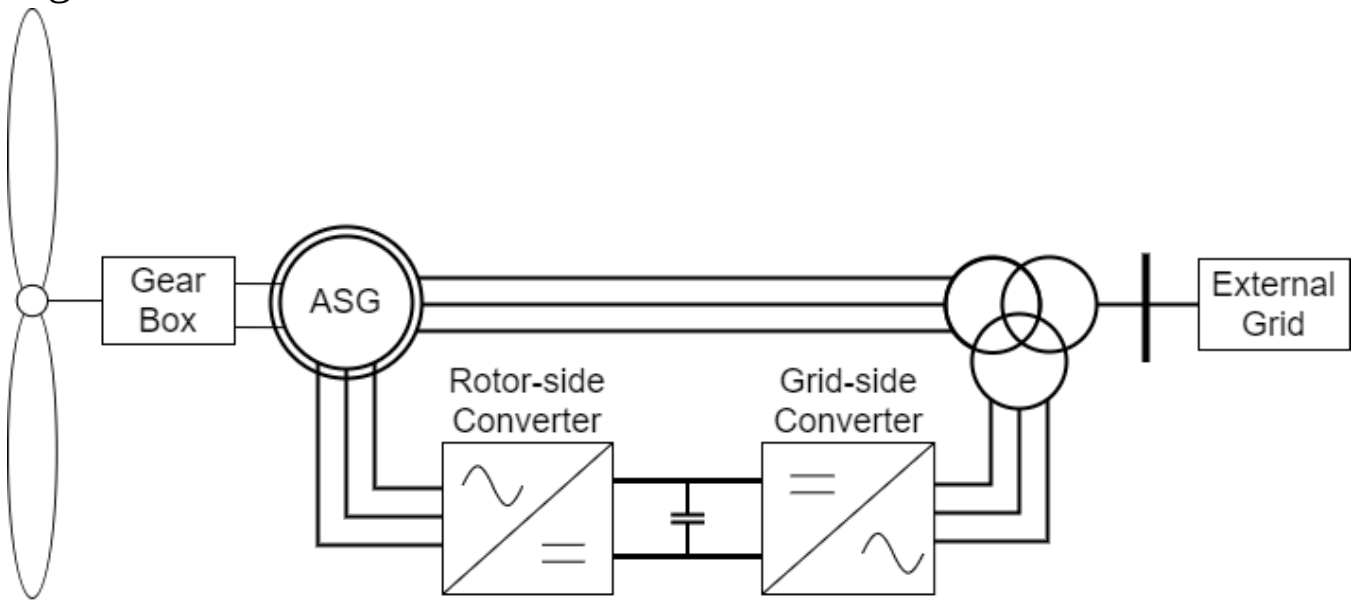

Fig. 1. DFIG model diagram.

Since the focus of this paper is the participation of DFIG in primary frequency regulation process, the electromagnetic transients of DFIG are neglected to decrease the computational effort of simulation and only the electromechanical transients are considered.

For each DFIG within a wind farm, the captured mechanical wind power at the local wind speed is given by the aerodynamic

model as

$$P_m = \frac{1}{2}\rho A_{rot} c_p(\lambda,\theta) v_m^3 \quad (1)$$

where $\rho$ is the air density, $A_{rot}$ is the cross-sectional area of rotational blade, $c_p$ is the wind power conversion coefficient, and $v_w$ is the current wind speed. $c_p$ is a nonlinear function of blade pitch angle $\theta$ and tip-speed ratio $\lambda$.

The definition of tip-speed ratio is given as

$$\lambda = \frac{\omega_r R}{v_w} \quad (2)$$

where $\omega_r$ and $R$ represent the rotation speed and the radius of wind turbine blade.

In practice, the relationship between $c_p$ and $(\lambda,\theta)$ is given by wind turbine manufacturers in the form of a set of curves or look-up tables. A typical power coefficient curve given by GE company is shown in Fig. 2, from which one can see the strong nonlinearity that couples the state variable $\omega_r$ between the input variable $v_w$.

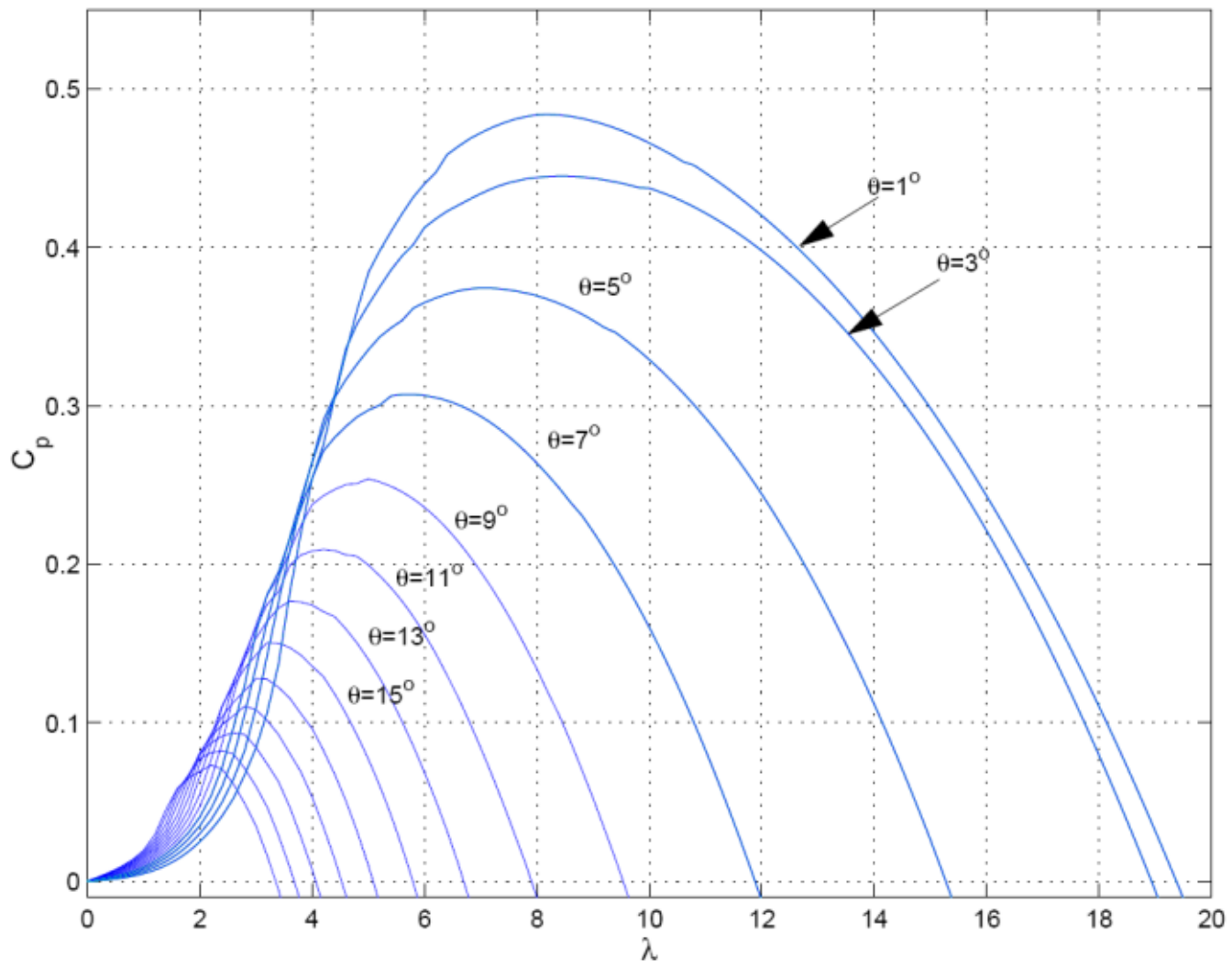

Fig. 2. Wind power coefficient curve [19].

The drive-train system containing the shaft and the gearbox is modelled as a two-mass model as follows:

$$2H_R \frac{d\omega_r}{dt} = t_m - K_s \delta_{rg} \quad (3)$$

$$2H_G \frac{d\omega_g}{dt} = K_s \delta_{rg} - t_e \quad (4)$$

$$\frac{d\delta_{rg}}{dt} = \omega_r - \omega_g \quad (5)$$

where $H_R$ and $H_G$ represent the inertia of the wind turbine and the generator, respectively, $t_m$ and $t_e$ are the mechanical and electromagnetic torque, $\omega_g$ is the speed of the generator rotor, the parameter $K_s$ represents the shat stiffness and the value $\delta_{rg}$ is the twist of the drive-train system.

The mechanical torque and the electromagnetic torque are defined separately as:

$$t_m = \frac{P_m}{\omega_r} \quad (6)$$

$$t_e = \frac{P_e}{\omega_g} \quad (7)$$

where $P_e$ is the active output power of DFIG.

Since the electromagnetic transients are neglected as mentioned, the electrical components including converters, transformers and asynchronous generator are modelled as a static generator that features like a controlled current source. The active and reactive power control of DFIG are separated through the decoupled control of d-axis and q-axis components of rotor current [20]. And for the coordinated control strategy, the reference value for d-axis current is calculated through the active power dispatch reference from the central wind farm controller and the terminal voltage of DFIG [15]. Due to the fast power regulation capacity of the back-to-back converters in DFIG, the active output power is considered to follow the dispatch reference of active power from the wind farm controller as:

$$P_e = P_{ref} \quad (8)$$

The final part that completes the dynamic modelling of DFIG is the pitch control system, which is integrated into the local controller of each DFIG for the protection of wind turbine from overspeed. The pitching process will only be activated when the current wind speed exceeds the nominal value. The demonstrated pitch control system is illustrated in Fig. 3, in which the time constant block represents the slow dynamic of the mechanical actuation process within DFIG during pitch control.

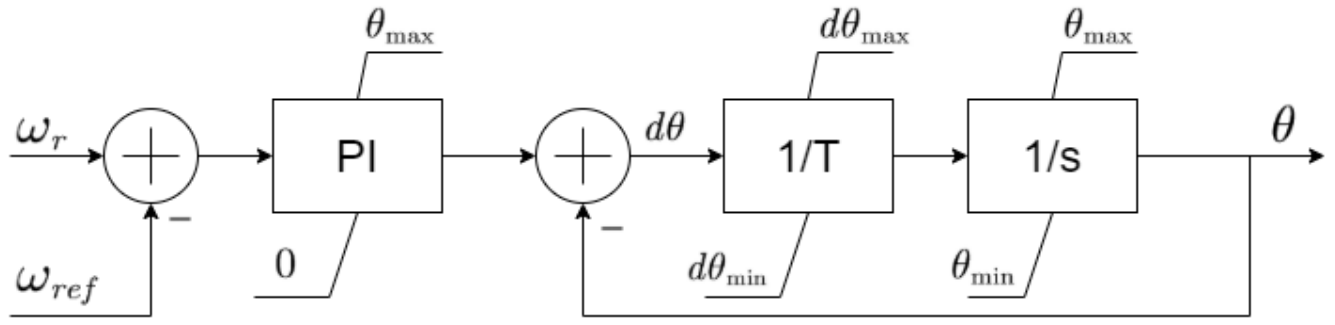

Fig. 3. Pitch control system

## III. WIND FARM CONTROLLER DESIGN

### A. Koopman Operator Theory

To make this paper more readable, we give the brief definition of our main tool Koopman operator on a discrete time autonomous dynamic system

$$x_{k+1} = f(x_k) \quad (9)$$

where $x \in \mathcal{M}$ is a n-dimensional state on a smooth manifold $\mathcal{M}$ and $f: \mathcal{M} \to \mathcal{M}$ is the nonlinear transition function.

And we define a set of real valued scalars $\psi: \mathcal{M} \to \mathbb{R}$ as the observable functions on the state space. Denote $\mathcal{F}$ as the function space of observables, $\psi \in \mathcal{F}$, we define the Koopman operator $\mathcal{K}: \mathcal{F} \to \mathcal{F}$ on the function space as follows:

$$\mathcal{K}\psi(x_k) = \psi(f(x_k)) \quad (10)$$

Worthy to mention, here we cast no requirement on the structure of $\mathcal{F}$. If $\mathcal{F}$ is a vector space, the Koopman operator becomes a linear operator. Further considering the dimension of $\mathcal{F}$ to be finite, then the Koopman operator is finite-dimensional and can be represented as a matrix [21]. Unfortunately, for most nonlinear dynamic system, it is hard to construct such a finite-dimensional observable space with a good reflection of the internal states, which makes the Koopman operator often referred to as an *infinite-dimensional linear operator*. Therefore, in practice, the Koopman operator on the real system is usually analyzed through its numeric approximation for the limited computational memory.

### B. *Specialized Dynamic Mode Decomposition* (*SDMD*)

This section introduces a SDMD algorithm as a data-driven approximation method of Koopman operator, which is related to the EDMD. When we focus on the frequency control of DFIG-based wind farm, the objective is to smooth the rotor speed fluctuation of DFIG during frequency response process, which specifies the dynamic control model of DFIG as

$$\omega_{r,k+1} = f(\omega_{r,k}, u_k) \tag{11}$$

where $\omega_{r,k}$ represents the rotor speed of DFIG at time $k$ and $u_k$ is the control input defined as

$$u_k = [P_{ref,k} \quad v_{w,k}]^\top \tag{12}$$

where $P_{ref,k}$ is the active power reference at time $k$ and $v_{w,k}$ is the local wind speed at time $k$.

And we define the extended state

$$\chi_k = \begin{bmatrix} \omega_{r,k} \\ \boldsymbol{u}_0 \end{bmatrix} \tag{13}$$

where $\chi_k \in \mathbb{R} \times \ell(\mathbb{R}^2)$ is the extended state, which is the product of the original state and all control sequences, and $\boldsymbol{u}_0 = (u_k)_{k=0}^{\infty} \in \ell(\mathbb{R}^2)$ is the control sequence.

The extended state inherits the nonlinear dynamics from the original state as

$$\chi_{k+1} = F(\chi_k) = \begin{bmatrix} f(\omega_{r,k}, \boldsymbol{u}_0(0)) \\ \boldsymbol{u}_1 \end{bmatrix} \tag{14}$$

where $\boldsymbol{u}_0(0)$ denotes the current control input and $\boldsymbol{u}_1$ denotes the left shifted control sequence. It can be found that the extended state space is infinite dimensional. However, the observable functions with the associated Koopman operator work the same way as in finite-dimensional uncontrolled dynamic system [18][22].

The modelling of DFIG based on SDMD method should start with the selection of observable functions on the extended state, which, as required by Koopman operator theory, should be coordinated with the inner nonlinearity of DFIG dynamics. Based on the experience-based model of DFIG in section II, we take some effort on settling the following observables

$$\boldsymbol{\psi}(\chi_k) = [\omega_{r,k} \quad \boldsymbol{\varphi}(\omega_{r,k}, u_k) \quad u_k]^\top \tag{15}$$

$$\boldsymbol{\varphi}(\omega_{r,k}, u_k) = \begin{bmatrix} \dfrac{1}{v_{w,k}} & \dfrac{\omega_{r,k}}{v_{w,k}} & e^{-\alpha \cdot \omega_{r,k} \cdot v_{w,k}} & \omega_{r,k}^2 \end{bmatrix}^\top \tag{16}$$

where state variable $\omega_{r,k}$ and input $u_k$ are reserved for linear control design and $\alpha$ is the coefficient generated from the equivalent formula of power coefficient curve $c_p(\lambda, \theta)$.

To approximate Koopman operator with finite dimensional matrix, a collection of data that records the evolving trajectory of the controlled dynamic system is expected. By generating several trajectories of (11) with random initial states, snapshot pairs $(\chi_k, \chi_{k+1})$ of current and future state are collected and expected to contain the enough information to reveal the dynamic behavior of the studied system. In the hope of approximating Koopman operator, we intend to find the matrix $\tilde{A}$ satisfying the following optimization problem

$$\min_{\tilde{A}} \sum_{k=1}^{N} \left\| \boldsymbol{\psi}(\chi_{k+1}) - \tilde{A}\boldsymbol{\psi}(\chi_k) \right\|_2^2 \tag{17}$$

where $N$ is the length of collected data.

For linear control design, the structure of observables $\boldsymbol{\psi}$ can be further reformed as

$$\boldsymbol{\psi}(\chi_k) = \boldsymbol{\psi}(\omega_{r,k}, u_k) = [z_k \quad u_k]^\top \tag{18}$$

where $z_k \in \mathbb{R}^{N_{lift}}$ represents the lifted state containing the monitored state and hyper-dimensional observables. The dimension $N_{lift}$ of lifted state in DFIG case is 5 specifically.

Since we care only for the impact of control input on the state rather than the evolving of the input variable, the transition matrix $\tilde{A}$ is partitioned as follows

$$\tilde{A} = \begin{bmatrix} A & B \\ C & D \end{bmatrix} \tag{19}$$

where $A$ is a square matrix of dimension $N_{lift} \times N_{lift}$ , $B$ is input matrix of dimension $N_{lift} \times 2$ and $C, D$ are irrelevant due to the focus on the control problem.

Subtract the calculation of $C, D$ in (17)

$$\min_{A,B} \sum_{k=1}^{N} \| z_{k+1} - Az_k - Bu_k \|_2^2 \tag{20}$$

The problem can be solved with its analytical solution based on the pseudoinverse calculation of the data matrix [22], which makes the online modelling of controlled system convenient. And the solution to (20) results in the following linearized hyper-dimensional dynamic model of the studied system

$$z_{k+1} = Az_k + Bu_k \tag{21}$$

The form in (21) is suitable for linear control design, e.g., MPC. Since the monitored state is listed at the front rows, the boundary constraint and objective function associated to the monitored state can be done by the manipulation of the first several elements of the lifted state.

Worthy to mention, the selection of observable functions in SDMD differs from the dictionary fashion in EDMD in that the prior knowledge of the controlled system, i.e., the experience-based model, is utilized to guide the trial-and-fail process during the selection, which powers the data-driven modelling method with the human understanding of the intrinsic dynamics of the studied system. Such *specialized* design of observables is proven in section IV with improvements of both simplicity and accuracy. The rationale behind is that when the nonlinearity of observable functions is coordinated with the intrinsic nonlinear dynamics of the studied system, it is expected to build an invariant subspace of the infinite dimensional Hilbert space $\mathcal{H}$, which enables the global linearization of the nonlinear dynamic system [23]. Review the result in (16), an explicit explanation can be found when we recall the major nonlinearity of DFIG.

The first and the second term correspond to the definition of tip speed ratio, which plays an important role in determining the captured wind power. The third term is extracted from the equivalent expression of power coefficient based on [10]. And the last term corresponds to the kinetic energy stored in the rotational mass of DFIG. To be mentioned, the result in (16) is not final due to the nonlinearity of finding the *best* combination of observables itself, which means the selection of observables is not a process of accumulation. Since the accuracy of SDMD relies on the invariance of the observable space, a good design of observables is always a set of nonlinear functions that corresponds well to the intrinsic nonlinearity of the studied system with finely tuned correlation between the observables themselves. However, to find such a functional observable design is expected to be easier when we have the guidance from the explicit form of experience-based model of the studied system. And the trial-and-failure process is worthy due to the dramatically decreased lifting dimension needed to linearize the dynamic system. Compared with EDMD, the prior effort in tuning the nonlinear observables trades off the limited online optimization time, which is more favorable in real fields.

### C. Wind Farm Frequency Control

During wind farm participating in frequency response, the active output power of wind farm is expected to follow the temporal power deviation order, which is calculated from the speed-droop characteristic required by the system operator, and the inner state, i.e., the rotor speed of each DFIG, is to be monitored and dynamically optimized to decrease its fluctuation level, which is likely to extend the lifetime of fragile parts including gears and shaft within DFIG for their mechanical fatigue is relieved due to the optimization control. Therefore, we give the following objective function for MPC

$$\min_{P_{ref}} \sum_{i=1}^{M} \sum_{k=0}^{T-1} \| \omega_{r,k+1}^{i} - \omega_{r,k}^{i} \|_2^2 \tag{22}$$

where $M$ is the number of controlled units and $T$ is the length of horizon of MPC.

For wind farm participating in primary frequency response, the active power output deviation constraint is given as

$$\sum_{k=1}^{M} \Delta P_{ref,k}^{i} = K_{df} \Delta f \tag{23}$$

where $K_{df}$ is the speed-droop radio, and $\Delta P_{ref,k}^{i}$, $\Delta f$ are defined as

$$\Delta P_{ref,k}^{i} = P_{ref,k}^{i} - P_{MPPT}^{i} \tag{24}$$

$$\Delta f = f_e - f_{ref} \tag{25}$$

where $P_{MPPT}^{i}$ represents the active power order from local controller under MPPT mode and $f_e$, $f_{ref}$ are frequency measured at gird connection point and the frequency reference.

With SDMD method described above, for each generation unit, which is either a single DFIG machine or an integrated DFIG group, e.g., a row of DFIGs with similar wind speed and mechanical parameters, the following dynamic model is expected to be available

$$\begin{bmatrix} \omega_{r,k+1}^{i} \\ \boldsymbol{\varphi}_{k+1}^{i} \end{bmatrix} = A_i \begin{bmatrix} \omega_{r,k}^{i} \\ \boldsymbol{\varphi}_{k}^{i} \end{bmatrix} + B_i u_k^i \tag{26}$$

For the safe operation of DFIGs, the boundary constraints on the rotor speed and the active power reference are given as

$$\omega_{\min} \leqslant \omega_{r,k}^{i} \leqslant \omega_{\max} \tag{27}$$

$$P_{\min} \leqslant P_{ref,k}^{i} \leqslant P_{\max} \tag{28}$$

Finally give the initial condition of hyper dimensional observables

$$\boldsymbol{\psi}(\omega_{r,0}^{i}, u_0^i) = [\omega_{r,0}^{i} \;\; \boldsymbol{\varphi}(\omega_{r,0}^{i}, u_0^i) \;\; u_0^i]^\top \tag{29}$$

And the equations from (22) to (29) form the complete MPC problem for wind farm frequency control. Since the dynamic model has been fully linearized, the fast and accurate solution to such a convex optimization problem is guaranteed.

The overall control structure is illustrated in Fig. 4.

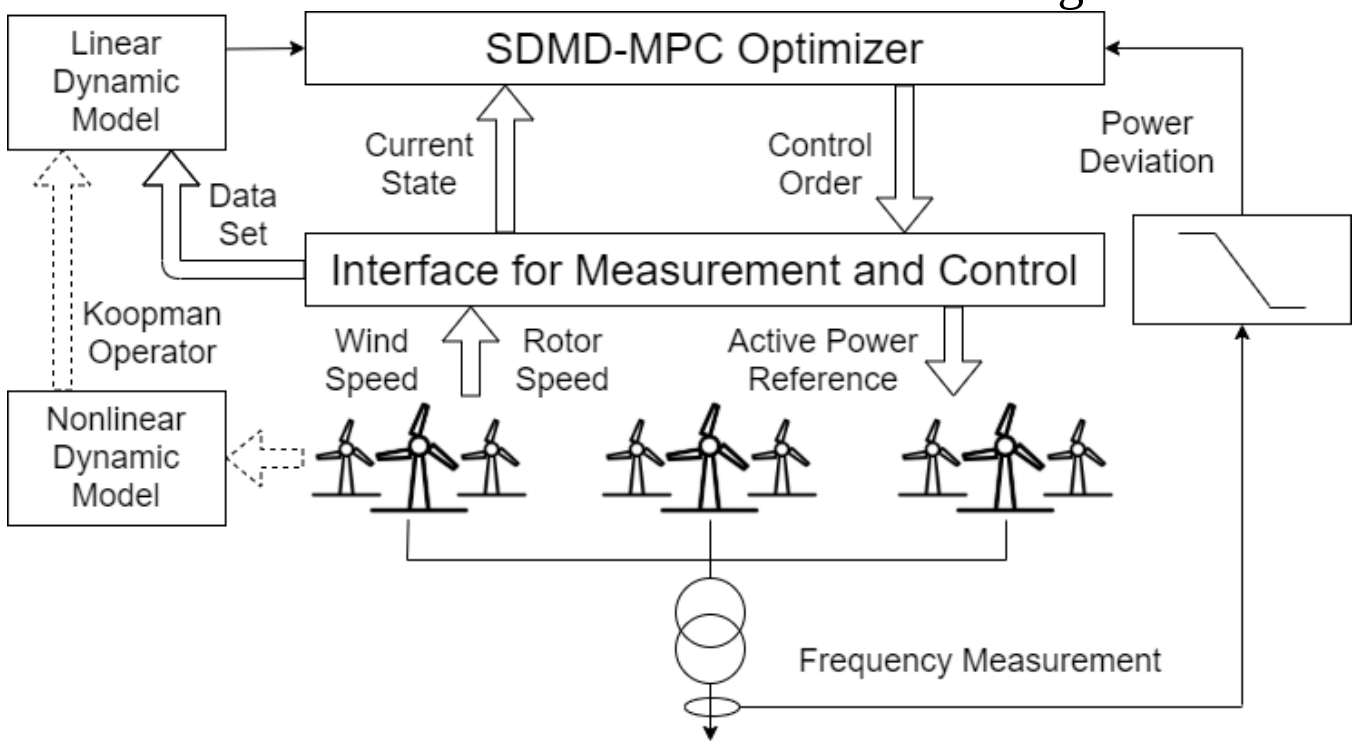

Fig. 4. Wind farm frequency controller

## IV. CASE STUDY

### A. Data-driven Modelling of DFIG

To verify the effectiveness of the proposed SDMD method, the dynamic simulation model of a wind farm integrated 4-bus system is implemented in DIgSILENT/PowerFactory. The diagram of the tested system is shown in Fig. 6. The rated apparent power of single DFIG is 1.5 MVA and the terminal voltage is 0.69 kV. The synchronous machine is set with hydro governor of IEEEG2 type. The wind farm is comprised of three DFIG groups. Each group contains 50 1.5 MW DFIG, which makes the total wind penetration reach 53%. The simulation step is set to 0.1s while the length of prediction horizon is 100, which means the performance of data-driven model is evaluated based on its dynamic behavior approximation for future 10s. As the benchmark, EDMD method is implemented with 100 radial basis functions as observables. The historical data set contains the past 10min online measurement of wind speed, rotor speed and active power reference, which is shown in Fig. 7. The dynamics of strong nonlinearity make the local linearization method hard to fit the data accurately in global horizon, and the piecewise linearization has difficulty in remaining a balance between accuracy and simplicity. Therefore, only Koopman operator related algorithm are compared based on 100 randomly generated scenarios, with the probability distribution of wind speed obeying Weibull distribution [24].

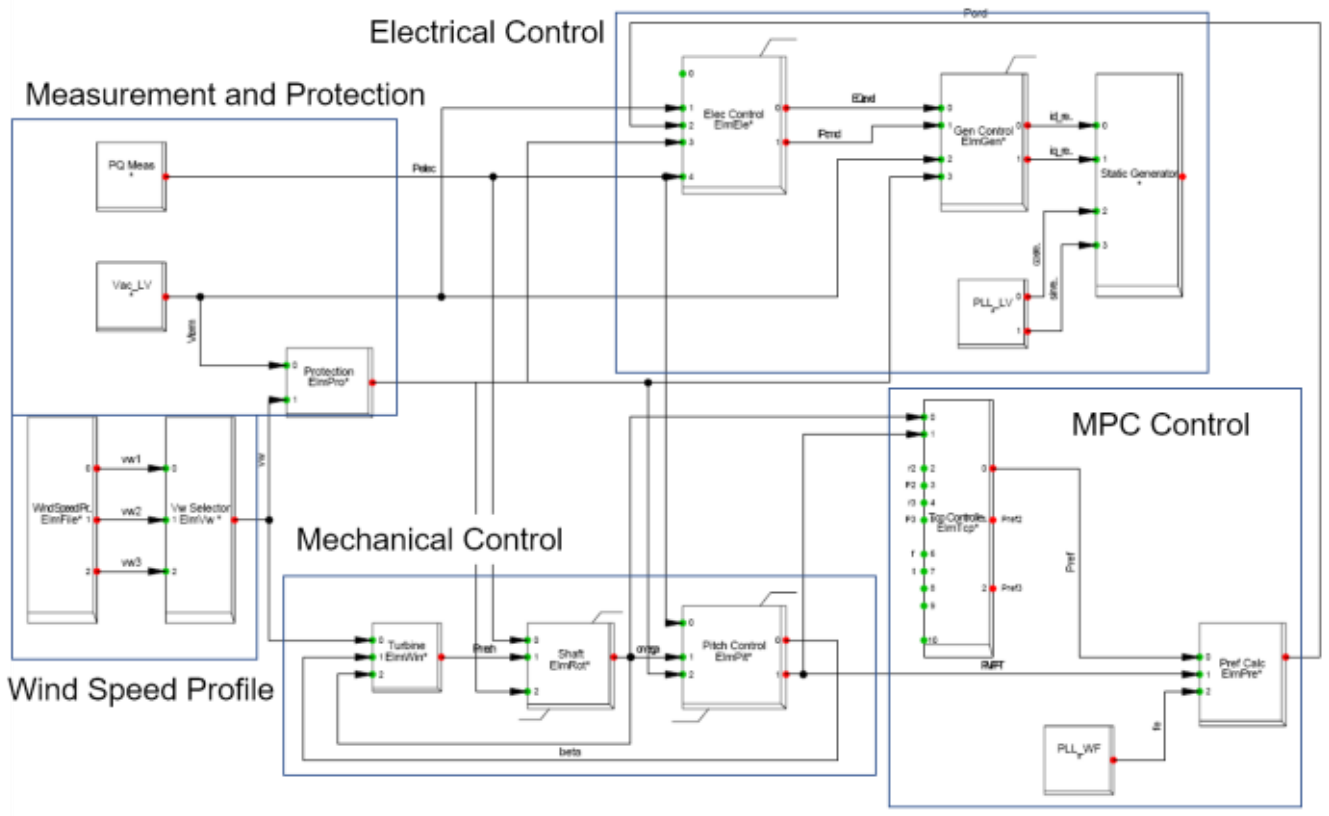

Fig. 5. Diagram of single DFIG simulation model.

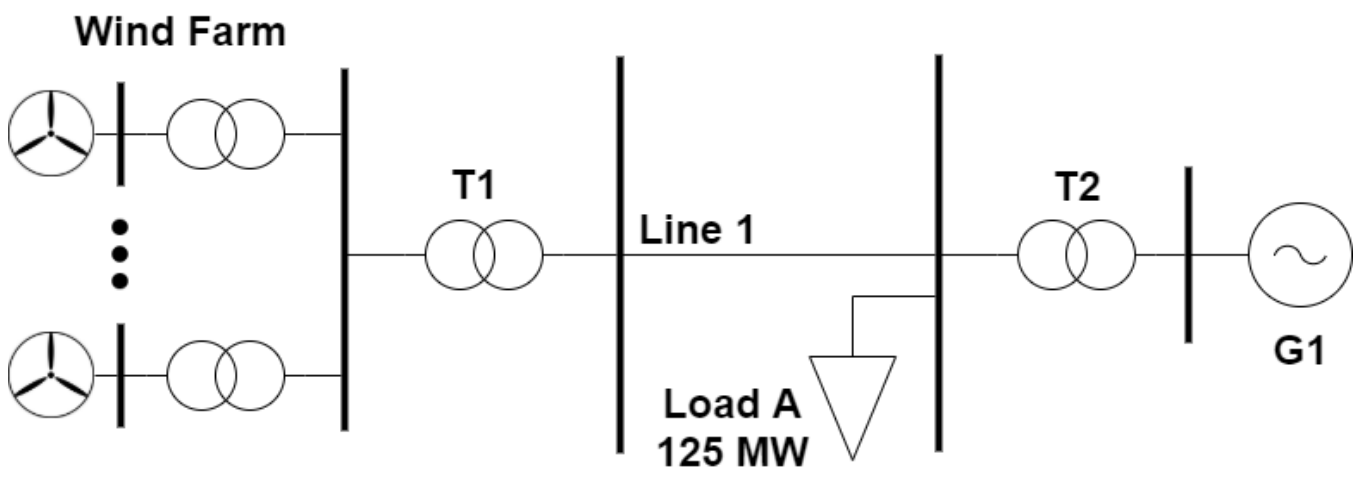

Fig. 6. Diagram of wind farm integrated power system

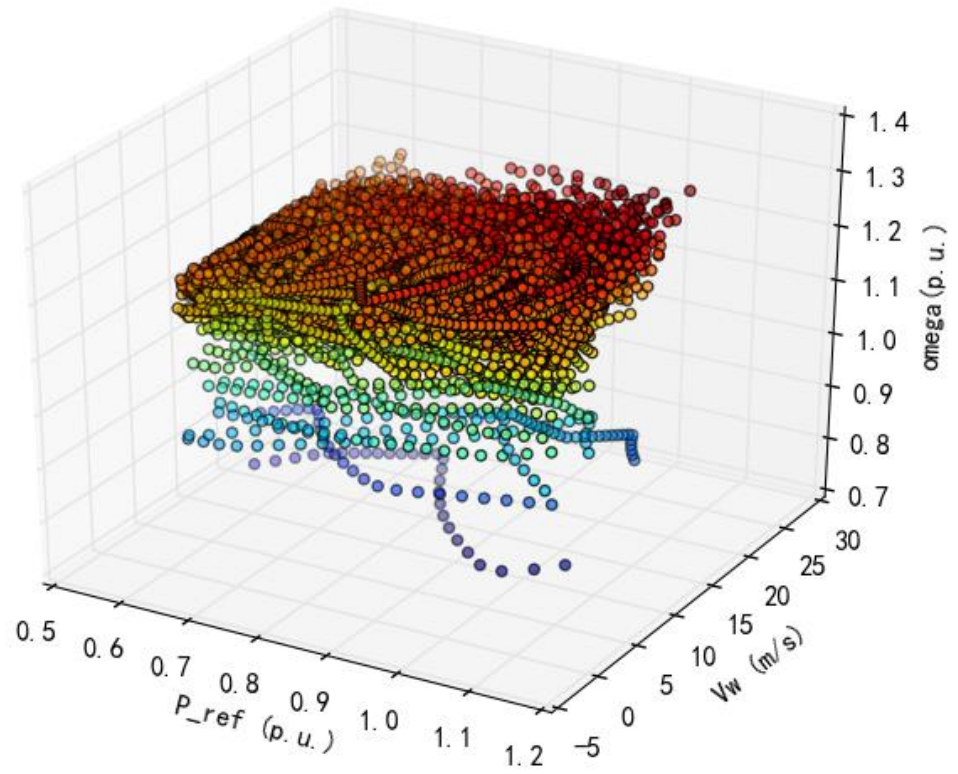

Fig. 7. Historical data set.

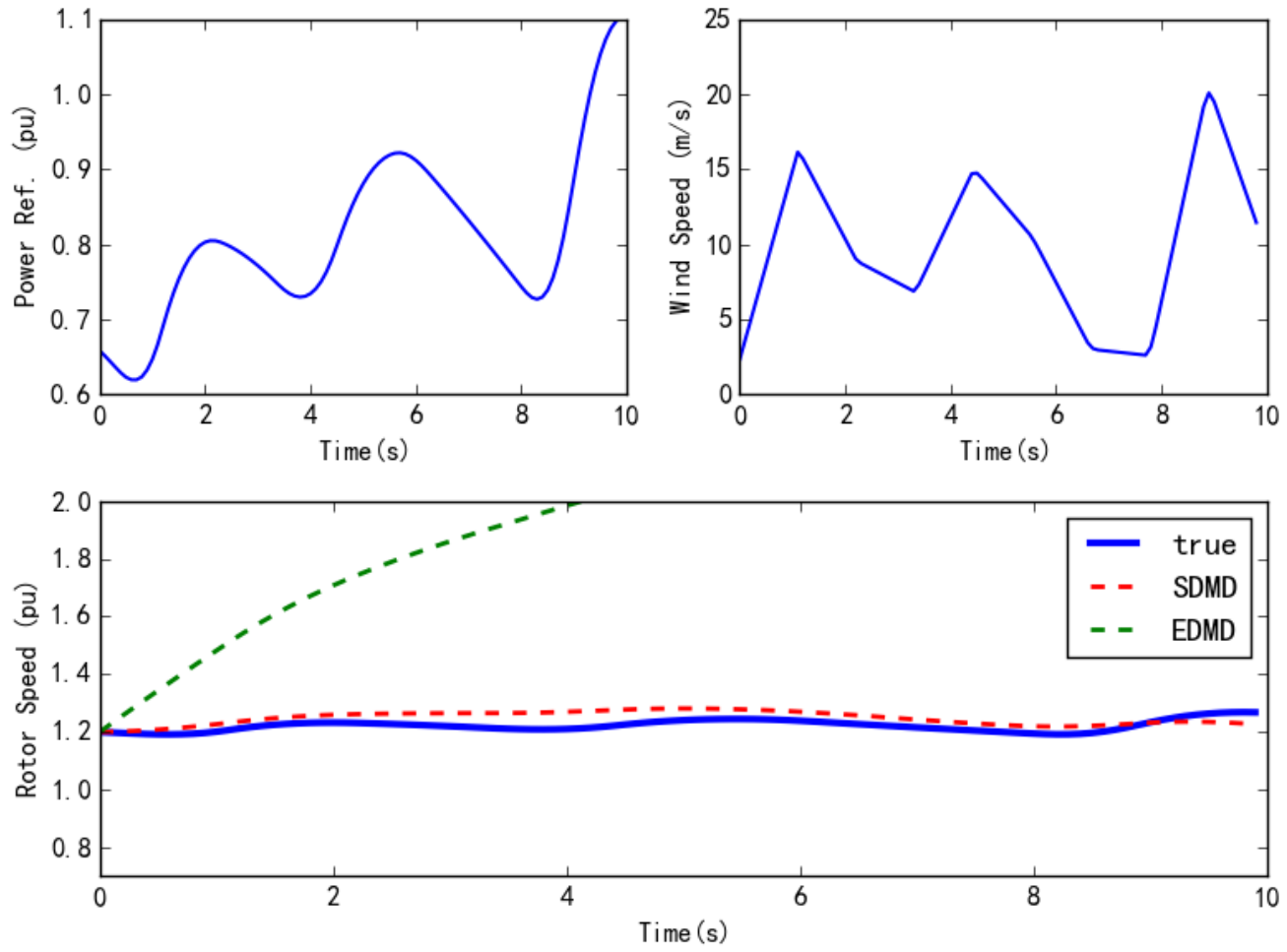

Fig. 8. Prediction comparison for DFIG

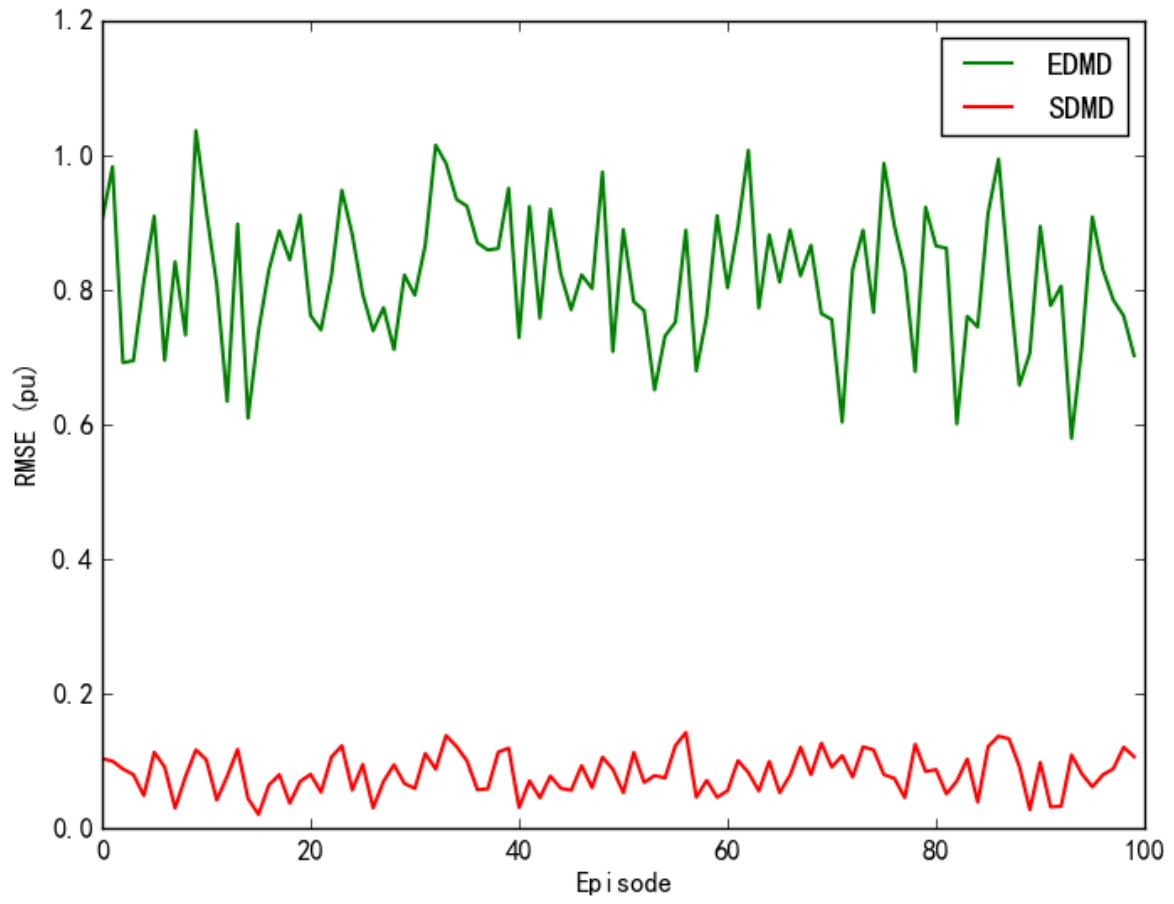

Fig. 9. Statistical prediction comparison for DFIG

TABLE I
STATISTICAL COMPARISON OF PREDICTION PERFORMANCE OF DATA-DRIVEN MODELLING METHODS FOR DFIG

| Method | *Dimension of Lifted State* | *Average RMSE (p.u.)* |
|---|---|---|
| SDMD | 5 | 0.082 |
| EDMD | 101 | 0.819 |

Fig. 8 shows the global linearization capability of the proposed method indicating that the specially selected observables in SDMD fit the dynamics of DFIG better than uniformed observables in EDMD. Further compare the prediction accuracy of two methods based on root mean squared errors (RMSE), which is defined between predicted rotor speed $\omega_{pred,k}$ and true rotor speed $\omega_{true.k}$ as

$$\mathrm{RMSE} = \sqrt{\sum_{k} (\omega_{pred,k} - \omega_{true,k})^{2}} \tag{30}$$

The comparison in Fig. 9 and Table I shows that the proposed method outweighs EDMD in average level with a stable performance for multiple scenarios. Furthermore, the SDMD method takes much less computational memory and runtime since the burden introduced by dimensional extending is relieved due to specialized design of observable functions. Such characteristics are favorable in online dynamic optimization control because the dynamic model generated is fully data-driven and linear with good accuracy for different working conditions.

## *B. Wind Farm Under-frequency Response*

To compare different control policies based on their frequency response capability, the local droop control and MPPT control are implemented with the wind speed faced by each group set as shown in Fig. 10. The frequency deviation is introduced through the load step change at 0.5s with a level up to 5%.

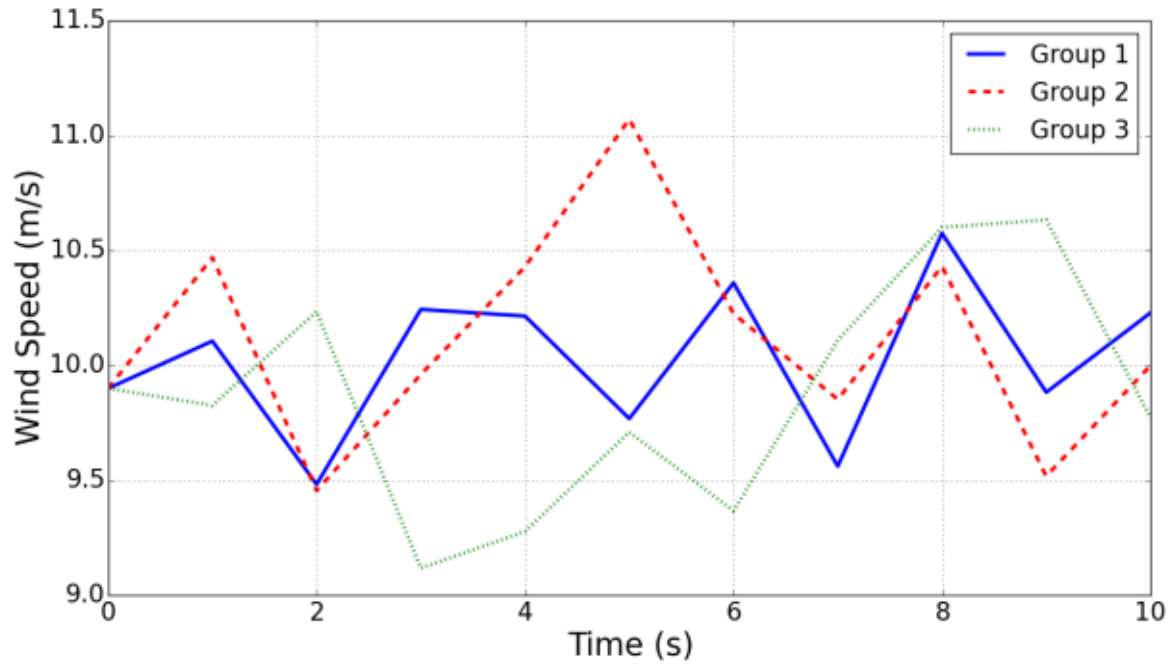

Fig. 10. Wind speed profile of three DFIG groups

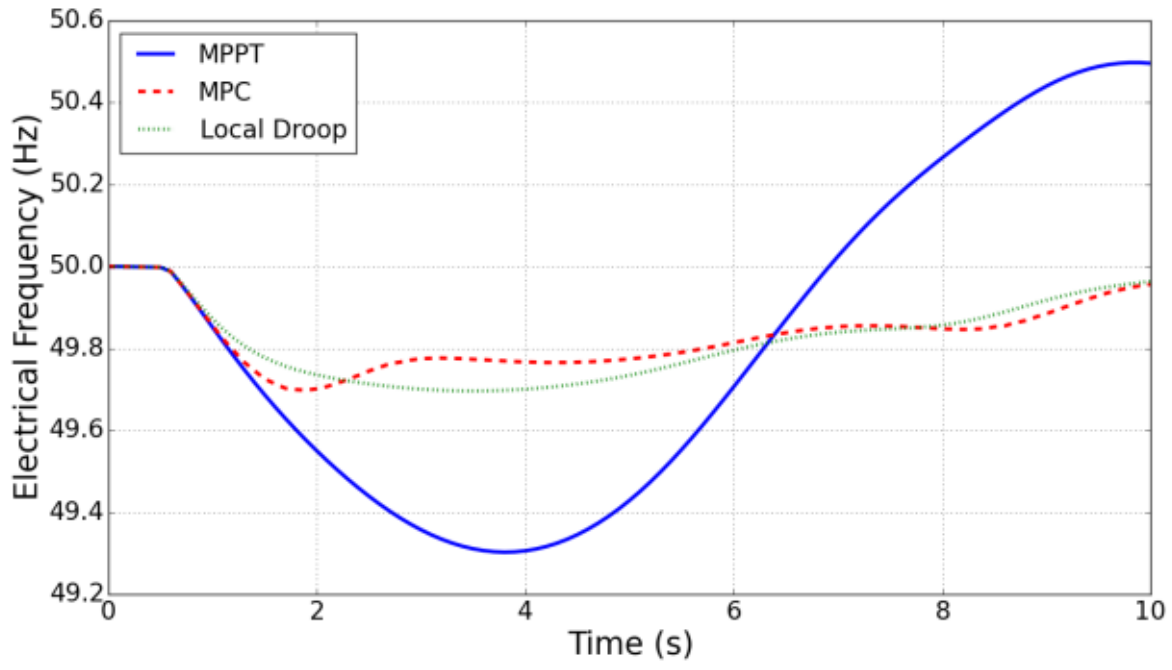

Fig. 11. Frequency dynamics of wind farm (Kdf=0.2 p.u./Hz).

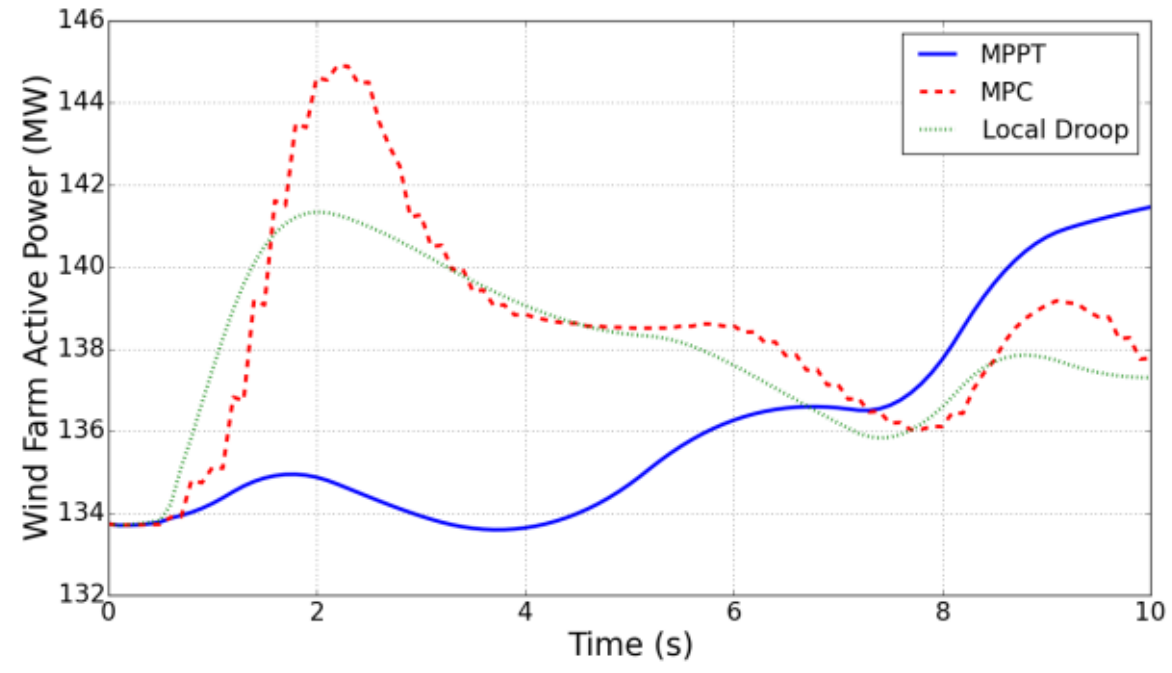

Fig. 12. Active power output of wind farm (Kdf=0.2 p.u./Hz).

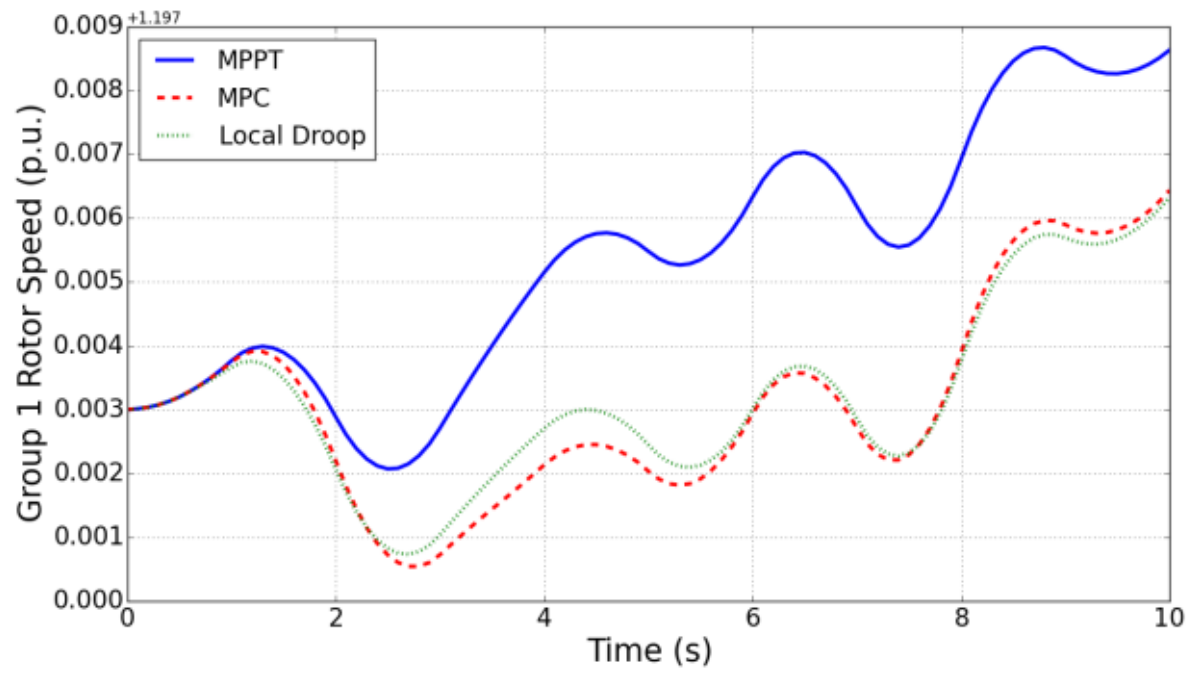

Fig. 13. Rotor speed of DFIG group 1 (Kdf=0.2 p.u./Hz).

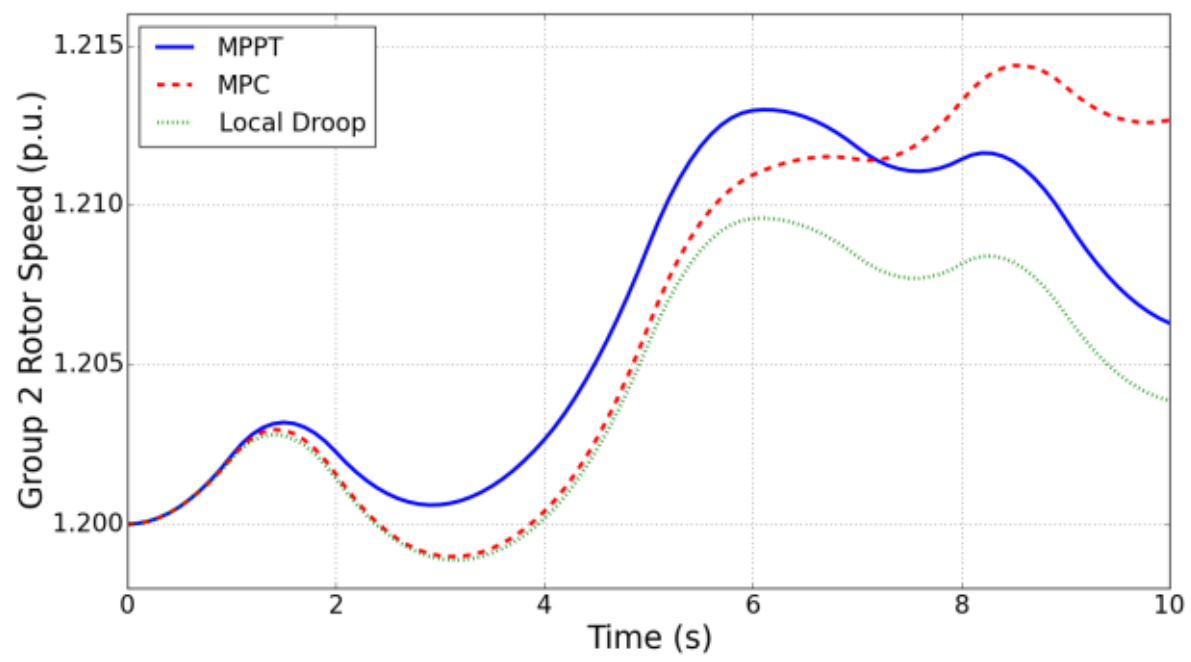

Fig. 14. Rotor speed of DFIG group 2 (Kdf=0.2 p.u./Hz).

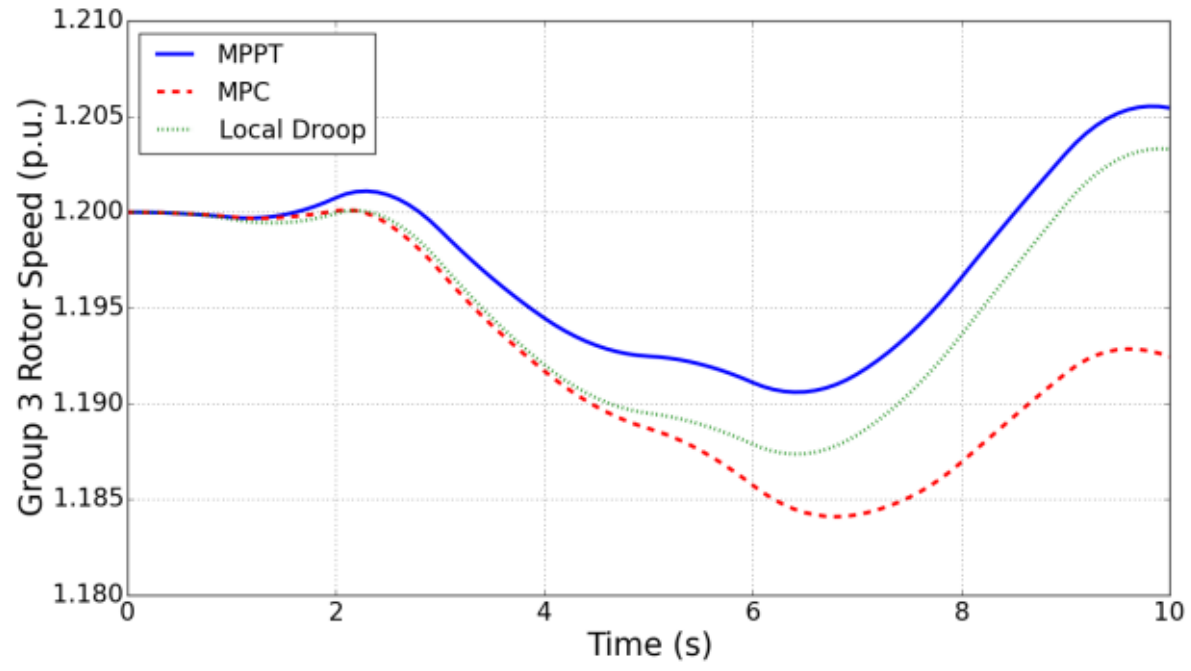

Fig. 15. Rotor speed of DFIG group 3 (Kdf=0.2 p.u./Hz).

The simulation results from Fig. 11 to Fig. 15 present the recording of the dynamic process of wind farm under-frequency response. Since the focus of this paper is the primary frequency regulation, the simulation time and the horizon scope of MPC are both set to 10s. The control step of MPC is 0.1s. The speed-droop ratio of MPC keeps the same with local droop at 0.2 p.u./Hz. Under MPPT mode, the wind farm provides no support for system frequency stability and the load step change causes the system frequency drop down to 49.3 Hz. When the wind farm participates in primary frequency under local droop mode, the frequency minimum is drawn back to 49.7 Hz. Under the proposed MPC policy, the frequency minimum is also regulated to 49.7 Hz, which means the proposed method fully meets the speed-droop requirement from the system operator. Apart from outside frequency support service, the inner states of wind farm are dynamically optimized under MPC policy to minimize the introduced rotor speed distortion. The average level of rotor speed distortion is defined as follows

$$\gamma = \frac{1}{M \cdot T} \sum_{i=1}^{M} \sum_{k=0}^{T-1} \frac{\omega_{r,k+1}^{i} - \omega_{r,k}^{i}}{T_s} \tag{31}$$

where $T_s$ is the control step of MPC.

TABLE II
AVERAGE LEVEL OF ROTOR SPEED DISTORTION DURING FREQUENCY RESPONSE PROCESS (KDF=0.2 P.U./HZ)

| Control Policy | Rotor Speed Distortion (p.u./s) | Percentage Increase (%) |
|---|---|---|
| MPPT | $2.25\times10^{-3}$ | 0 |
| MPC | $2.16\times10^{-3}$ | -4.00 |
| Local Droop | $2.29\times10^{-3}$ | 1.78 |

Table II shows the comparison of rotor speed distortion during frequency response process. Since the active power reference follows the simple droop characteristic under local droop mode, the frequency regulation service causes an increase of 1.78% percentage. But under the proposed MPC method, the rotor speed dynamics are further stabilized even compared with the situation under MPPT mode, which means the disturbance from wind speed fluctuation is optimized as well, indicating the capability of MPC policy to optimize the dynamic behavior of wind farm and to participate in primary frequency regulation simultaneously. Worthy to mention, here we reserve no margin in wind farm active power output for primary frequency response and only temporary frequency support are activated. When the local control mode of each DFIG is switched to deloading mode, the MPC policy can provide better frequency support in long term with only slight adjustment on method design, e.g., extending the optimization horizon of MPC. Further considering the fully data-driven modelling and control structure adopted in proposed method, the effectiveness of optimization control is more convincing compared with model-based control policy.

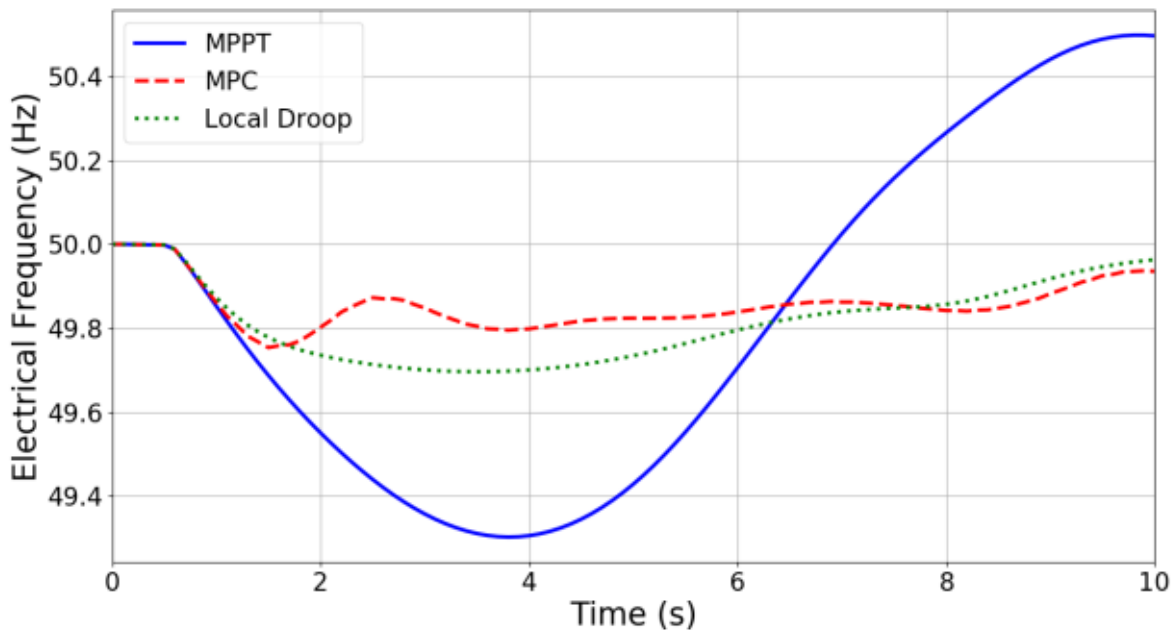

Fig. 16. Frequency dynamics of wind farm (Kdf=0.3 p.u./Hz).

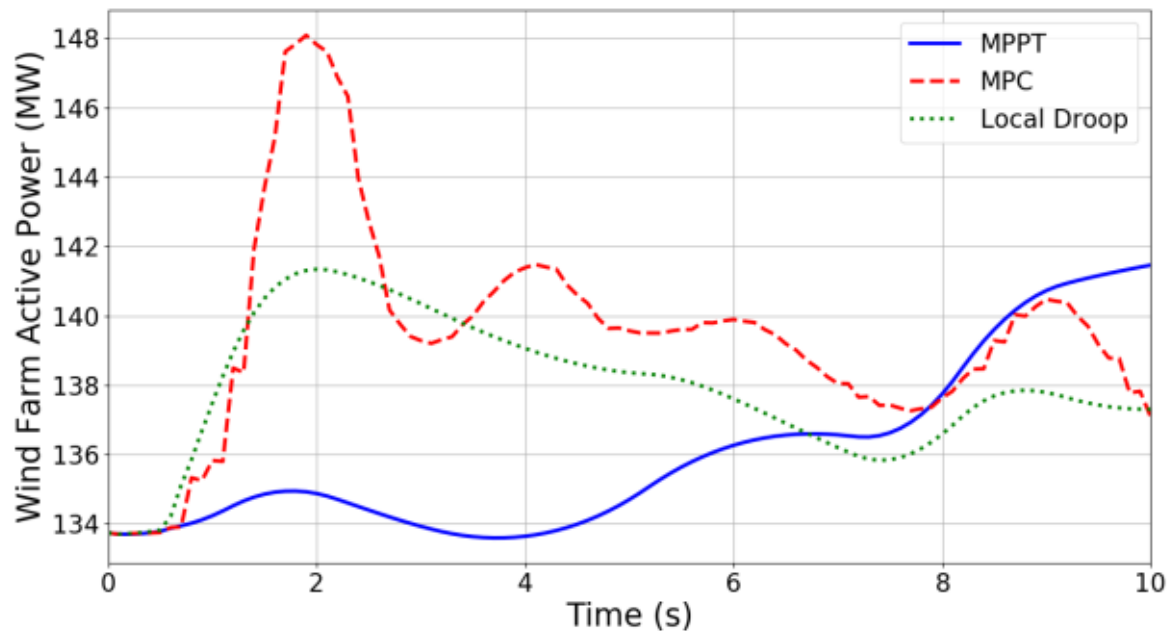

Fig. 17. Active power output of wind farm (Kdf=0.3 p.u./Hz).

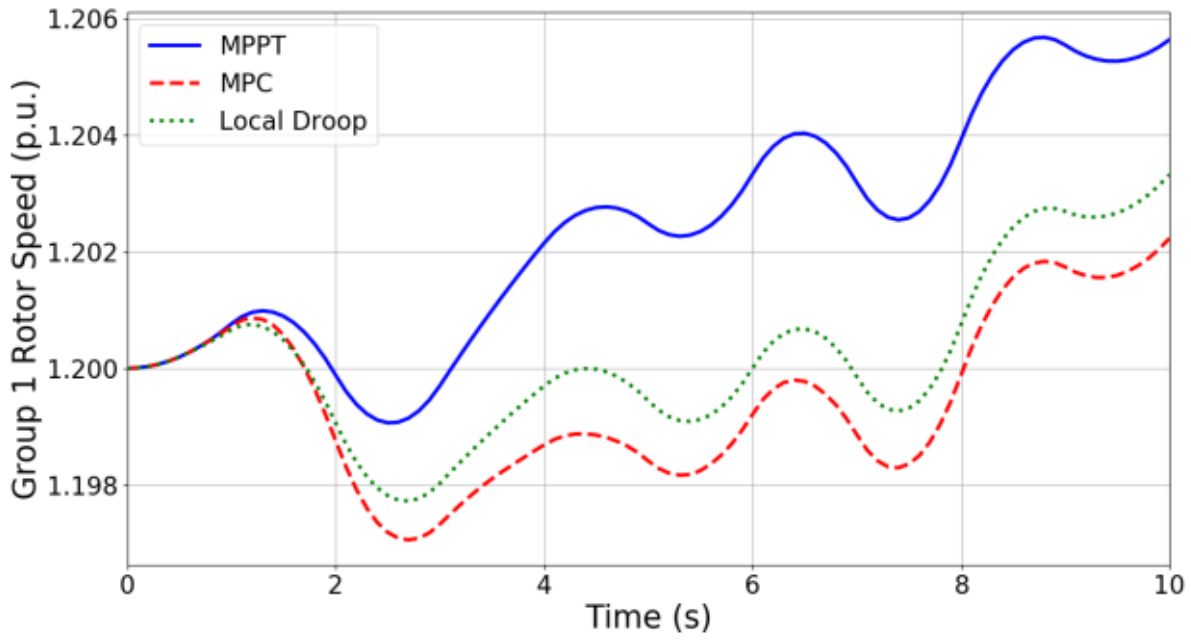

Fig. 18. Rotor speed of DFIG group 1 (Kdf=0.3 p.u./Hz).

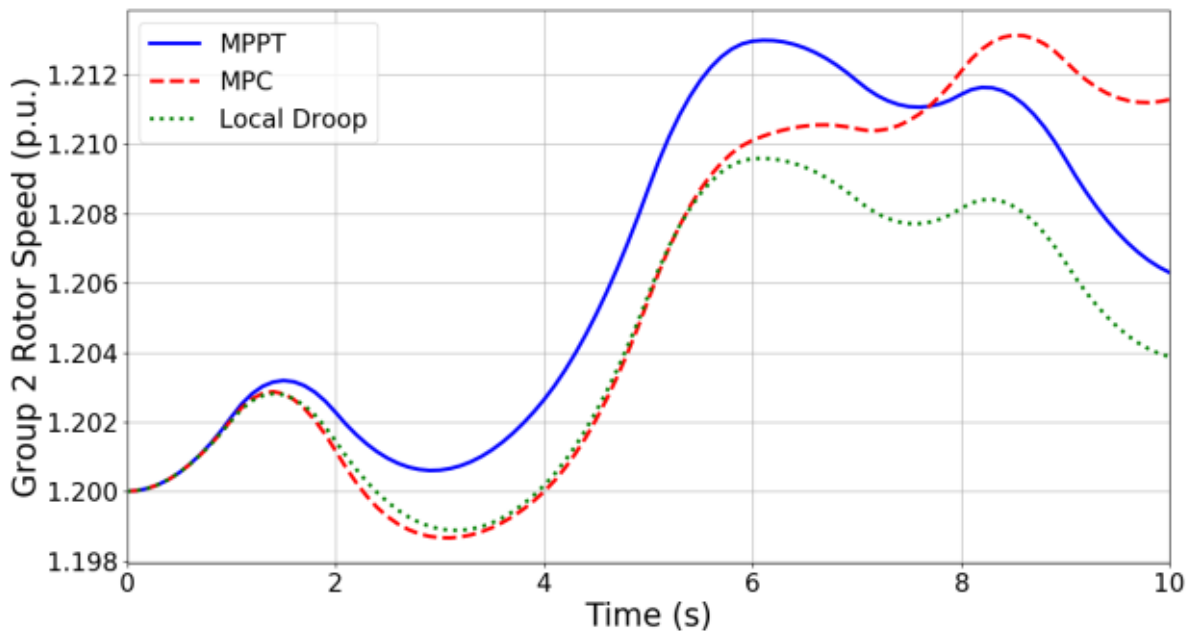

Fig. 19. Rotor speed of DFIG group 2 (Kdf=0.3 p.u./Hz).

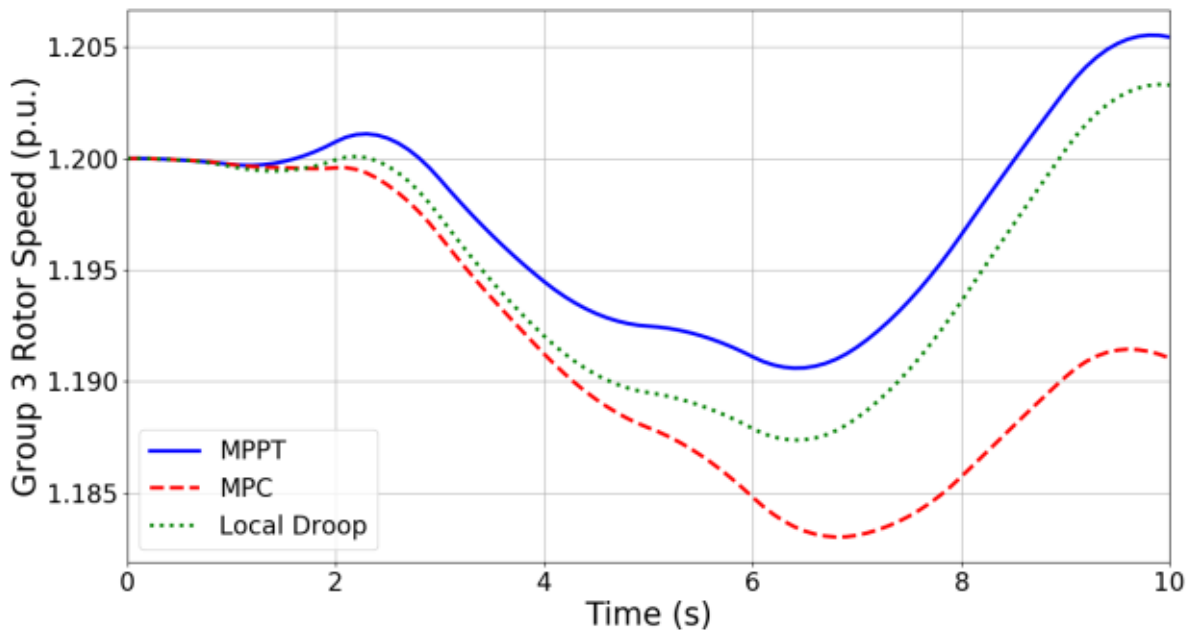

Fig. 20. Rotor speed of DFIG group 3 (Kdf=0.3 p.u./Hz).

### C. *Flexible droop characteristics*

Since the MPC policy only has to regulate central control parameter to adjust the speed-droop characteristics of the whole wind farm, the fast adjustment ability provides the system operator with flexible arrangement for primary frequency regulation service. Considering this, we tune up the speed-droop ratio of wind farm and run the simulation again. The recorded dynamic control process from Fig. 16 to Fig. 20 shows that the participation of wind farm in frequency regulation is deepened due to high speed-droop ratio. As a result, the minimum system frequency after the load step change is regulated to 49.75 Hz. Table III further analyzes the rotor speed dynamics during frequency response process. One can see that the rotor speed distortion is even more smoothed after tuning up the speed-droop ratio. However, it is observed that rotor speed of three DFIG groups is deviated from the original value deeper as the expense of frequency regulation. This rotor speed deviation issue can be better compensated when deloading control is adopted on the local side controller of each DFIG.

TABLE III
AVERAGE LEVEL OF ROTOR SPEED DISTORTION DURING FREQUENCY RESPONSE PROCESS (KDF=0.3 P.U./HZ)

| Control Policy | Rotor Speed Distortion (p.u./s) | Percentage Increase (%) |
|---|---|---|
| MPPT | $2.25\times10^{-3}$ | 0 |
| MPC | $2.15\times10^{-3}$ | $-4.44$ |
| Local Droop | $2.29\times10^{-3}$ | 1.78 |

## V. CONCLUSION

In this paper, a SDMD method is proposed based on Koopman operator theory for data-driven dynamic modelling of nonlinear system. Compared with existing EDMD method, the proposed method holds the advantage both in model accuracy and simplicity due to specialized design of observable functions, which is powered by the experience-based model of the studied system. Compared with local linearization and piecewise linearization method, SDMD inherits the merit of global linearization with the ability to capture the intrinsic nonlinearity of studied system. Applying this modelling method to DFIG, we construct a fully data-driven online optimization method for wind farm to participate in primary frequency regulation service. The case study demonstrates the effectiveness of proposed method on balancing the outside frequency response and inside rotor speed stability. And the proposed method also features flexible adjustment to the speed-droop characteristic, which enables wind farm to support system frequency stability more flexibly within security limit.